\documentclass[prb,aps,preprint]{revtex4}

\usepackage{graphicx}
\usepackage{dcolumn}
\usepackage{bm}

\newcommand{\I}{\mathrm{i}}
\newcommand{\rhou}{\rho_{\uparrow}}
\newcommand{\rhod}{\rho_{\downarrow}}

\newcommand{\rhos}{\rho_{S}}

\newcommand{\rhodutil}{\tilde{\rho}_{\downarrow \uparrow}}
\newcommand{\tilderhou}{\tilde{\rho}_{\uparrow}}
\newcommand{\tilderhod}{\tilde{\rho}_{\downarrow}}

\newcommand{\tilderhosu}{\tilde{\rho}_{S \uparrow}}
\newcommand{\tilderhosd}{\tilde{\rho}_{S \downarrow}}

\newcommand{\rhodutildt}{\dot{\tilde{\rho}}_{\downarrow \uparrow}}
\newcommand{\tilderhoudt}{\dot{\tilde{\rho}}_{\uparrow}}
\newcommand{\tilderhoddt}{\dot{\tilde{\rho}}_{\downarrow}}

\newcommand{\tilderhosdt}{\dot{\tilde{\rho}}_{S}}
\newcommand{\tilderhosudt}{\dot{\tilde{\rho}}_{S \uparrow}}
\newcommand{\tilderhosddt}{\dot{\tilde{\rho}}_{S \downarrow}}

\newcommand{\Wdu}{W_{\downarrow \uparrow}}
\newcommand{\Wud}{W_{\uparrow \downarrow}}
\newcommand{\Wsu}{W_{S \uparrow}}
\newcommand{\Wus}{W_{\uparrow S}}
\newcommand{\Wsd}{W_{S \downarrow}}
\newcommand{\Wds}{W_{\downarrow S}}
\newcommand{\Vdu}{V_{\downarrow \uparrow}}
\newcommand{\Vsu}{V_{S \uparrow}}
\newcommand{\Vsd}{V_{S \downarrow}}
\newcommand{\Wd}{W_{\downarrow}}
\newcommand{\Wu}{W_{\uparrow}}
\newcommand{\Ws}{W_{S}}

\newcommand{\spinup}{|\!\!\uparrow\rangle}
\newcommand{\spindown}{|\!\!\downarrow\rangle}
\newcommand{\singlet}{|S\rangle}

\newcommand{\spinupd}{\langle\uparrow\!\!|}
\newcommand{\spindownd}{\langle\downarrow\!\!|}

\begin{document}

\title{Electron spin tomography through counting statistics: a quantum 
trajectory approach}
\author{Holger Schaefers}
\email{holger.schaefers@physik.uni-freiburg.de}
\author{Walter T. Strunz}
\affiliation{Theoretische Quantendynamik, Physikalisches Institut,
Universit\"at Freiburg, Hermann-Herder-Stra{\ss}e 3, 79104 Freiburg,
Germany}

\begin{abstract}
We investigate the dynamics of electron spin qubits in quantum dots. Measurement of the qubit state is realized by a charge current through the dot.
The dynamics is described in the framework of the quantum trajectory 
approach, widely used in quantum optics, and we show that it can be applied
successfully to problems in condensed matter physics.
The relevant master equation dynamics is unravelled to simulate 
stochastic tunneling events of the current through the dot. 
Quantum trajectories are then used to extract the counting statistics of the current. We show how, in combination with an electron spin resonance (ESR) field, 
counting statistics can be employed for quantum state tomography of 
the qubit state. Further, it is shown how decoherence and relaxation 
time scales can be estimated with the help of counting statistics, in the time domain. Finally, we discuss a setup for single shot measurement of the qubit state
without the need for spin-polarized leads.
\end{abstract}


\maketitle

\section{Introduction}

Controlling and preserving coherent quantum dynamics 
in the framework of quantum information processing is a 
challenging task.\cite{Nielsen} Very recently,
more and more experiments on implementing such ideas in mesoscopic systems
based on solid state devices \cite{Leggett} have been realized, e.g.\ 
Josephson junctions,\cite{Vion,Heij,Astafiev} and also single electron 
spins in single defect centers.\cite{Jelezko}
The electron spin in quantum dots has been recognized early as a potential 
carrier of quantum information,\cite{Loss} but experimental developments 
of suitable mesoscopic devices have only recently been pursued.

In previous work it was shown how
quantum dots may serve as spin filters, or memory devices for electron 
spin.\cite{Recher} Important progress was made in both theoretical
and experimental research focusing on measurement schemes through charge 
currents.\cite{Engel1,Engel2,Engel3,Vandersypen,Elzerman1,Hanson1,Hanson2,Elzerman2} 
Even a single-shot readout of the electron spin state has been realized 
\cite{Elzerman3} and allows for the measurement of the relaxation time of 
a single spin. Still, the decoherence time of a single electron spin in a 
quantum dot has not yet been determined experimentally.

In some of these experiments important quantities are the counting 
statistics of tunneling electrons.\cite{Bagrets} As for charge qubits, 
a measurement on the single electron level may be achieved through a single 
electron transistor (SET) device \cite{Shnirman,Lu} or with a quantum 
point contact (QPC) close to the quantum dot.\cite{Vandersypen2, Elzerman3}

It is important to realize that the measurement through a charge current
itself has dynamical implications for the measured qubit. We are thus lead
to the problem of noise and statistics induced by the measurement process
in these mesoscopic systems.\cite{Makhlin}
Such problems have been tackled some time ago very elegantly through
the concept of {\sl quantum trajectories} in quantum optical applications.\cite{Carmichael, Plenio}
In particular, jump processes to describe
the time evolution of open systems while counting emitted quanta are
well established in the framework of systems that are described
by a master equation of Lindblad type. Such ideas have already been
applied to measurement processes based on quantum point contacts
in mesoscopic devices.\cite{Ruskov, Goan} 
In the context of quantum information processing, 
such quantum trajectory methods turn out to be essential for 
the design of active quantum error correcting codes \cite{alber} and, 
more generally, of quantum feedback mechanisms.\cite{wiseman}

Historically, a major driving force behind the development of quantum
trajectory methods were the growing possibilities to experiment with
single quantum systems in traps or cavities. More recently, such
experiments have been extended to mesoscopic solid state devices.
Therefore, we expect a growing need for such methods in these fields.

The aim of this paper is to show how quantum 
trajectories serve as a useful framework to discuss the physics of
mesoscopic carriers of quantum information under continuous measurement.
In particular, we determine counting statistics of electrons tunneling 
through a quantum dot, depending on the electron spin-state. 
We show how a simple setup for state
tomography can be achieved through a measurement of counting statistics
in combination with a coherent ESR field.
We display how decoherence and relaxation time scales can be extracted
from the measured data, in the time domain.

\section{Electron spin dynamics on the quantum dot}

\begin{figure}
\includegraphics[angle=0,width=0.3\textwidth]{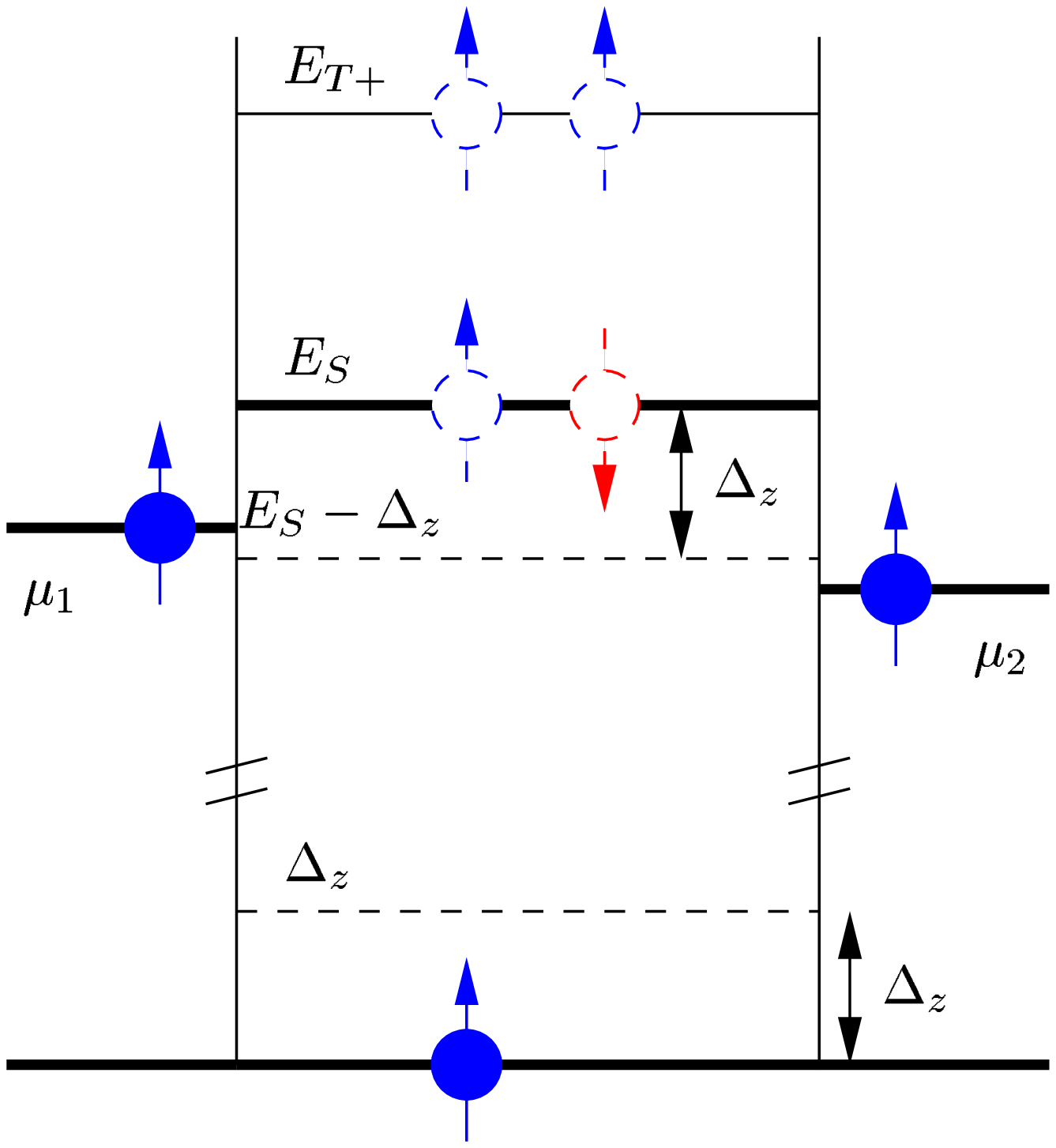}
\hspace*{0.2\textwidth}
\includegraphics[angle=0,width=0.3\textwidth]{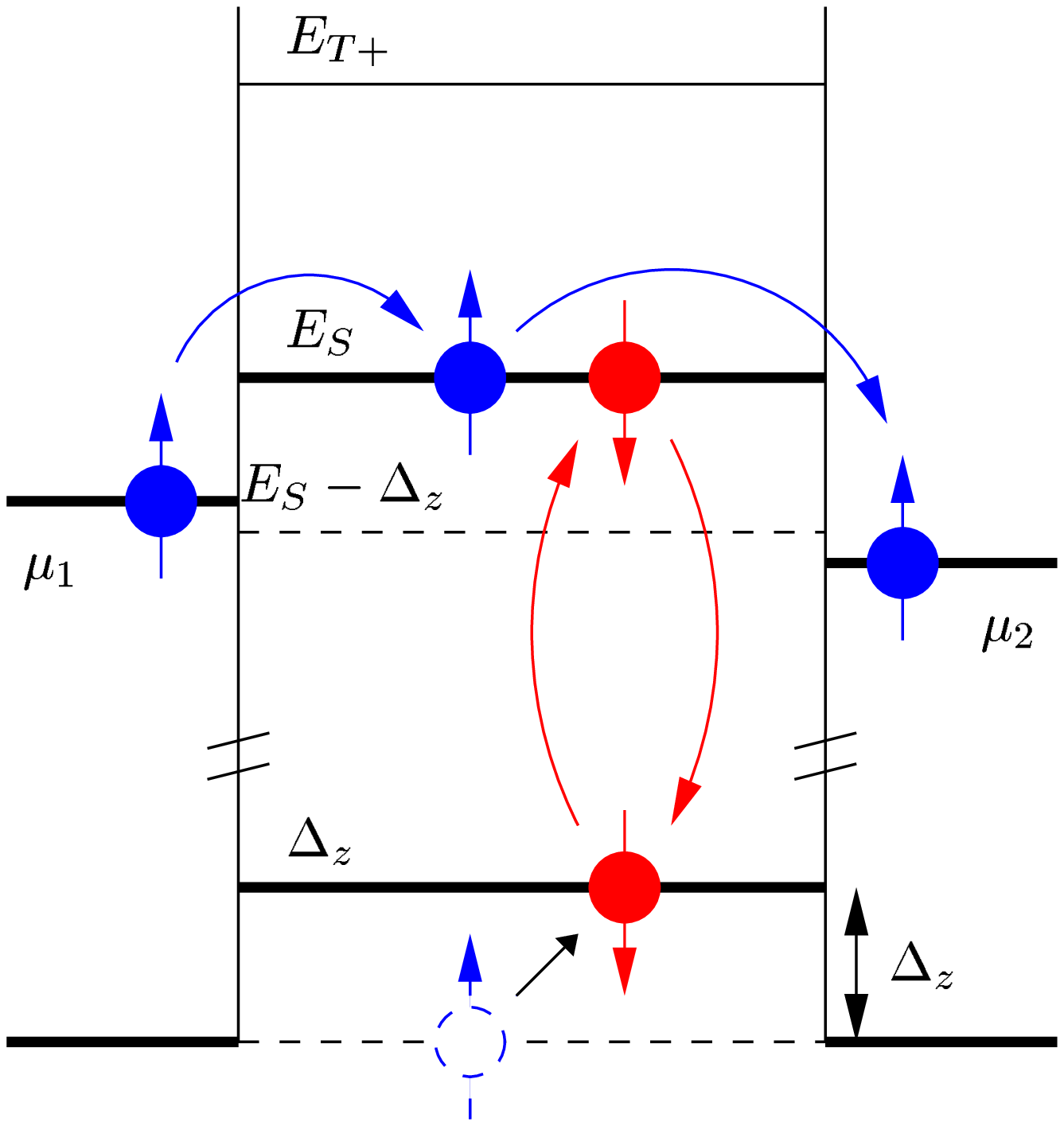}
\caption{\label{qdot}Closed dot (left): 
Chemical potentials are too small to allow an 
electron to tunnel onto the dot.
Open dot (right): After excitation of the dot electron, the chemical
potential $\mu_1$ is large enough for
an electron of lead one to tunnel onto the dot and form the singlet 
state with the dot electron.}
\end{figure}

We consider a quantum dot with spin-$\frac{1}{2}$ ground state 
in the Coulomb blockade regime as in
\cite{Recher, Engel1, Engel2}, see also Fig.\ref{qdot}.
The quantum dot is subject to a constant magnetic field $B_z$ which 
leads to a Zeeman splitting $\Delta_z=g \mu_B B_z$ of the electronic states,
where $g$ is the electron g factor and $\mu_B$ the Bohr magneton
(throughout this paper we use $|g|=0.44$ for GaAs and units 
such that $\hbar=1$). 
Two leads at chemical potentials $\mu_1$ and $\mu_2$ are coupled to the dot
for charge transport. Further, as in Ref.\ \onlinecite{Engel2},
we allow for an electron spin resonance (ESR) field to drive coherent
transition between the two spin states.

Leaving sources of uncontrollable environmental influences aside
for a moment (see below),
the total Hamiltonian consists of contributions from 
electrons on the dot, electrons in the leads and
a tunneling interaction between dot and leads,
\begin{equation}\label{htot}
  H_{\rm tot}=H_{\rm dot}+H_{\rm leads}+H_{\rm T}.
\end{equation}
Here, $H_{\rm dot}=H_0 + H_{\rm ESR}(t)$ contains contributions from
charging and interaction energies of the electrons on the dot, the 
interaction energy $-\frac{1}{2}\Delta_z \sigma_z$ with the
static magnetic field, and the
ESR-Hamiltonian $H_{\rm ESR}(t)=-\frac{1}{2}g \mu_B B(t) \sigma_x$
of the interaction of the electron spin with a magnetic field 
$B(t)=B^0_x \cos(\omega t - \varphi)$, 
oscillating linearly in the $x$ direction. 
The $\sigma_i$ $(i=x,y,z)$ denote the usual Pauli spin matrices.
The Hamiltonian for the two leads $(k=1,2)$ reads
$H_{\rm leads}=\sum_{kn\sigma}\epsilon_{kn} c_{kn\sigma}^\dagger c_{kn\sigma}$,
with $c^\dagger_{kn\sigma}$ the creation operator of an electron with orbital 
state $n$, spin $\sigma$ and energy $\epsilon_{kn}$ in lead $k$.
Finally, the coupling between dot and leads is described by the standard 
tunneling Hamiltonian 
$H_{\rm T}=\sum_{knm\sigma} T_{kn}^\sigma c_{kn\sigma}^\dagger d_{m\sigma}+$
h.c.\,, where we denote with $T_{kn}^\sigma$ a tunneling amplitude and 
with $d_{m\sigma}$ the annihilation operator of an electron on the dot 
in orbital state $m$.
Following Ref.\ \onlinecite{Engel2},
for the description of the dot dynamics in 
Sect.\ \ref{mastereq}
we will also include further (microscopically unspecified)
dissipative interactions between the dot states and their environment
that are not among the known contributions to the total
energy as they appear in Eq.\ (\ref{htot}).

In the following we give a qualitative picture of the 
relevant dot states,
see Fig.\ref{qdot}, more details may be found in Ref.\ \onlinecite{Engel2}.
For simplicity we assume there is only one electron on the dot. 
With $\sigma_z\spinup = +\spinup$, 
the electron has the ground state $\spinup$ with energy $E_\uparrow =0$ and the exited state $\spindown$ with $E_\downarrow = \Delta_z$. 
If an electron tunnels onto the dot, the two electrons can form the singlet 
state $\singlet=(|\!\!\uparrow\downarrow\rangle-|\!\!\downarrow\uparrow\rangle)
/\sqrt{2}$ with energy $E_S$
or either of the three triplet states. As the triplet state
$|T_+\rangle= |\!\!\uparrow\uparrow\rangle$ has higher energy 
(for suitable magnetic field \cite{Hanson2}),
the singlet $\singlet$ is the ground state for two electrons on the dot. 
The chemical potentials are chosen such that
$E_S > \mu_1=E_S -\Delta_z /2 > E_S -\Delta_z > \mu_2=E_S -3\Delta_z /2$. Under these conditions the dot 
can be opened and closed for a sequential tunneling current by 
a spin flip induced by an ESR field:\cite{Engel2}
An electron at chemical potential $\mu_1$ in the left lead and a dot 
electron in state $\spinup$ do not have sufficient 
energy to form the singlet state. If due to an ESR induced excitation 
the dot state is $\spindown$, however, less energy is required 
and an electron in lead one can tunnel onto the dot to form
the singlet state. Tunneling onto the dot from lead two 
is suppressed by several orders of magnitude if the thermal energy is much 
lower than the energy gap even if the dot electron is in 
the excited state $\spindown$. 
At higher temperatures, to assure that the singlet can only be formed with 
the excited dot electron, one can choose spin-polarized leads. This may be
achieved with several methods, see Ref.\ \onlinecite{Engel2} and references therein.

Thus, within these constraints we see that the current in lead one 
is proportional to the probability for the dot being in the excited state,
i.e.\ $I_1^\uparrow (t)\propto \rhod$, while the current in lead two is 
proportional to the probability of the dot being in the singlet state, 
$I_2^\uparrow (t)\propto \rhos$, see Ref.\ \onlinecite{Engel2}.

\subsection{Master equation}\label{mastereq}

The traditional description of the dynamics of the dot is based
on a master equation for the reduced density operator of the dot,
obtained from the total density matrix $\rho_{\rm tot}(t)$ by
tracing over the degrees of freedom  of the leads:
$\rho_{\rm dot}=Tr_{\rm leads}[\rho_{\rm tot}]$.
As usual, we denote matrix elements with
$\rho_{nm}=\langle n|\rho_{\rm dot}|m\rangle$ (or 
$\rho_{n}=\langle n|\rho_{\rm dot}|n\rangle$) and include only the three 
relevant dot states $n,m \in \{ \uparrow,\downarrow, S\}$. 
We assume the dot and the leads 
to be uncorrelated initially, 
$\rho_{\rm tot}(0)=\rho_{\rm dot}(0) \otimes \rho_{\rm leads}(0)$. 
Starting from the von Neumann equation for the full density operator 
${\dot\rho}_{\rm tot}=
-\I\left[ H_{\rm tot},\rho_{\rm tot}\right]$, the master 
equation for $\rho_{\rm dot}$ was derived in Ref.\ \onlinecite{Engel2} using standard
methods within the Markov approximation. 
Further, we will allow for an arbitrary (fixed) phase
$\varphi$ of the ESR field which will play an important role
in determining the spin state. 

In order to eliminate the explicit time dependence emerging from the
ESR field, we here base our analysis on the dot state in a rotating frame,
\begin{equation}\label{rotframe}
\tilde{\rho}_{\rm dot}(t) \equiv 
e^{\I\omega t|\downarrow \rangle\langle\downarrow|} \rho_{\rm dot}(t)
e^{-\I\omega t|\downarrow \rangle\langle\downarrow|}.
\end{equation}
In fact, with the exception of
$\tilde\rho_{\downarrow\uparrow} = e^{i\omega t}\rho_{\downarrow\uparrow}$,
$\tilde\rho_{\downarrow S} = e^{i\omega t}\rho_{\downarrow S}$ and the
corresponding transposed expressions,
this transformation leaves almost all matrix elements
untouched.
 
Along the lines of the derivation in Ref.\ \onlinecite{Engel2},
one finds for the dot state in the rotating frame (\ref{rotframe})
a master
equation of {\sc Lindblad} form.\cite{Lindblad} It can be written as 
\begin{equation}\label{masterlind}
\partial_t{\tilde\rho_{dot}}
=\mathcal{L}\tilde\rho_{dot}\equiv -\I [ H,\tilde\rho_{dot} ] 
+ \frac{1}{2}\sum_{nm}( [ L_{nm} \tilde\rho_{dot} , L_{nm}^\dagger ] 
+ [ L_{nm} , \tilde\rho_{dot} L_{nm}^\dagger ] )
\end{equation}
with the time-independent Hamiltonian (in rotating wave approximation)
\begin{equation}\label{RWAH}
  H=(\Delta_z-\omega) \spindown\langle\downarrow |+E_S \singlet\langle S|
-\frac{\Delta_x}{4}(e^{-\I\varphi}\spinup\langle \downarrow |
+e^{\I\varphi}\spindown\langle \uparrow |)
\end{equation}
and the 
operators $L_{nm}=\sqrt{W_{nm}}|n\rangle\langle m|$ describing
incoherent transitions between levels $m$ and $n$
with a rate $W_{nm}$.

In particular, the four operators $L_{S \uparrow}, L_{S \downarrow},
L_{\uparrow S}$ and $L_{\downarrow S}$ describe transitions from
and to the singlet state and hence correspond to the tunelling of
an electron off or onto the dot. These four contributions give rise to
the current in the leads and are derived from the underlying
Hamiltonian (\ref{htot}). 
The rates are 
$\Wsd=\Wsd^1 + \Wsd^2$ with $\Wsd^l=\gamma_l^\uparrow f_l(E_S-\Delta_z)$ and 
$\Wds=\Wds^1 + \Wds^2$ with $\Wds^l=\gamma_l^\uparrow (1-f_l(E_S-\Delta_z))$ 
where $f_l(E)=\left[1+e^{(E-\mu_l)/kT}\right]^{-1}$ is the Fermi function of 
lead $l$. Analogously we define the rates $\Wsu$, $\Wus$, $\Wsu^l$, $\Wus^l$ 
with $\gamma_l^\downarrow$ and $f_l(E_S)$. Here 
$\gamma_l^\uparrow= 2\pi \nu_l^\uparrow |T_l^\uparrow|^2$ and 
$\gamma_l^\downarrow= 2\pi \nu_l^\downarrow |T_l^\downarrow|^2$ are the 
transition rates with density of states $\nu_l^{\uparrow,\downarrow}$ and 
tunneling amplitude $T_l^{\uparrow,\downarrow}$.\cite{Engel2} In the limit 
$kT \ll \Delta_z$ we have $\Wsd \approx \gamma_1^\uparrow$ and 
$\Wds \approx \gamma_2^\uparrow$, which resembles the sequential tunneling 
from lead one onto the dot and into lead two. Furthermore, we have $\Wsu \approx 0$ and 
$\Wus \approx \gamma_1^\downarrow + \gamma_2^\downarrow$, because we 
choose $\mu_1 , \mu_2 < E_S$. Throughout this paper we assume equal rates 
for both leads $\gamma_1^\uparrow=\gamma_2^\uparrow=\gamma^\uparrow$ and 
$\gamma_1^\downarrow=\gamma_2^\downarrow=\gamma^\downarrow$. Finally,
we set $\gamma=\gamma^\uparrow=\gamma^\downarrow$ if the leads are not 
spin-polarized, and $\gamma=\gamma^\uparrow$, $\gamma^\downarrow=0$ in
the case of spin-polarization.

We simulate stochastically all processes that could be observed in
principle, but eventually extract the desired information from
those quantities that correspond to the specific measurement scheme chosen.
Quantum transitions between the dot states
may be observed by monitoring the current through the dot, 
which is the starting point for our quantum trajectory analysis of the 
following Sections.

By contrast, mechanisms for incoherent 
spin flips (described by the operators
$L_{\downarrow \uparrow}$ and $L_{\uparrow \downarrow}$) and
dephasing mechanisms (described by the projectors 
$L_n\equiv L_{nn}=\sqrt{W_n}|n\rangle\langle n|$) are
introduced on phenomenological grounds and not contained in the
Hamiltonian (\ref{htot}). The (phenomenological) spin flip rates
are assumed to satisfy the condition of detailed balance:
$\Wud / \Wdu = e^{\Delta_z/k_B T}$. 
The rates $W_n$ are phenomenological 
dephasing rates:
the effect of an operator $L_{nn}$ in (\ref{masterlind})
is to kill coherences between state $|n\rangle$ and
the remaining states (at a rate $W_n$), while leaving 
probabilities unaffected.

If the coupling to the leads is switched off
(by an appropriate choice of the chemical potentials), the dynamics
as described by the master equation (\ref{masterlind}) is that
of a standard decaying two-state spin-system. Then the corresponding
(intrinsic) relaxation and decoherence rates turn out to be
\begin{eqnarray}\label{t1t2}
1/T_1 & = & \Wdu+\Wud \\ \nonumber
1/T_2 & = & (1/T_1+\Wu+\Wd)/2.
\end{eqnarray}

let us now turn to the dot dynamics: in terms of its coefficients,
the time evolution of the dot state $\tilde\rho_{\rm dot}$ given by the 
master equation (\ref{masterlind}) reads
\begin{eqnarray}
\tilderhoudt &=&-\frac{\Delta_x}{2}  {\rm Im}(e^{-\I\varphi}\rhodutil)
-(\Wdu + \Wsu)\tilde\rhou + \Wud \tilde\rhod  + \Wus \tilde\rhos\label{rwa1},
\\
\tilderhoddt &=&\frac{\Delta_x}{2} {\rm Im}(e^{-\I\varphi}\rhodutil)
-(\Wud + \Wsd)\tilde\rhod + \Wdu \tilde\rhou + \Wds \tilde\rhos \label{rwa2},
\\
\tilderhosdt &=&-(\Wus + \Wds)\tilde\rhos + \Wsu \tilde\rhou + \Wsd \tilde\rhod\label{rwa3},
\\
\rhodutildt&=&-(\I (\Delta_z-\omega) + \Vdu)\rhodutil + 
\I \frac{\Delta_x}{4} e^{\I\varphi} 
(\tilderhou - \tilderhod).    \label{rwa4} \\
\tilderhosudt &=&-(\I E_s + \Vsu)\tilderhosu
-\I \frac{\Delta_x}{4} e^{\I\varphi} \tilderhosd, \label{rwa5}
\\ 
\tilderhosddt &=&-(\I (E_s-\Delta_z+\omega) + \Vsd)\tilderhosd
-\I \frac{\Delta_x}{4} e^{-\I\varphi} \tilderhosu, \label{rwa6}
\end{eqnarray}
with the effective rates
\begin{eqnarray}
\Vdu&=&\frac{1}{2}(\Wdu+\Wud+\Wsu+\Wsd+\Wd+\Wu)=\frac{1}{2}(\Wsu+\Wsd)
+\frac{1}{T_2} , \;\;{}
\\
\Vsu&=&\frac{1}{2}(\Wdu+\Wsu+\Wus+\Wds+\Ws+\Wu),
\\
\Vsd&=&\frac{1}{2}(\Wud+\Wsd+\Wds+\Wus+\Ws+\Wd).
\end{eqnarray}
Note that Eqs.\ (\ref{rwa5}, \ref{rwa6}) are decoupled from 
Eqs.\ (\ref{rwa1}-\ref{rwa4}), and the latter are the only ones of
relevance to us. They enable us to determine easily counting statistics
of tunneling electrons numerically by 
means of the quantum trajectory method which we describe in the following 
Sections.

\section{Quantum Trajectories}

One major motivation behind the development of
quantum trajectory methods were experiments with
with single quanta. Before these developments,
naturally, ensemble experiments required simple ensemble theories.
Matters changed with experiments involving single atoms, electrons or 
ions in traps.
Continuously monitoring those systems, single quantum jumps
became visible to the bare eye. A theory of
continuous quantum measurement taking into
account continuous measurement records of the observed environment to 
update the 
quantum state accordingly, were developed, mainly with an eye on 
applications in quantum optics.

Experiments on the single quantum level have reached solid state
devices, as for instance electrons in quantum dots. Accordingly, the dynamics
of such nano-scale quantum systems may be described adequately by
quantum trajectories. In fact, it may well turn out that these
methods are even more useful in solid state devices since the
sensitivity of electron detectors is typically far better than that 
for photon detectors, on the single quantum level.

As will be explained in the following,
a quantum trajectory $\rho_c(t)$ describes a subensemble of the full 
(ensemble)
density operator $\rho(t)$, {\it conditioned} on a certain 
(stochastic) measurement
record, here detection events at certain times.
In this approach we determine the dynamics of an
electron on a quantum dot, conditioned on the measured 
(stochastic) tunneling current through the dot.

Quantum trajectory methods have changed remarkably the way we think about
open quantum system dynamics. While traditionally an open quantum
system is described by its density operator $\rho(t)$
as in the last section, quantum trajectories describe open system dynamics
taking into account certain continuous, stochastic
measurement outcomes. In other words, with quantum trajectories
one determines a {\it conditioned} density operator $\rho_c(t)$,
reflecting knowledge obtained from a continuous monitoring of the
environment.
Sampling over all these possible measurement 
records, in other words, ignoring the state of the environment,
one recovers the usual full ensemble $\rho(t)$. We write
$\rho(t) = {\cal M}[\rho_c(t)]$ where ${\cal M}[\ldots]$ denotes
the ensemble mean over all possible measurement records with
corresponding probability (see below).

The principle idea is to monitor the environment rather than
ignoring, i.e.\  tracing over it. In quantum optics one tries
to detect photons emitted from the quantum system of interest,
here we detect electrons in the leads coupled to the quantum dot.

In order to illustrate this approach, we consider a simplified open
quantum system -- the generalization to the quantum dot case will be
obvious. This model system consists of two levels and is coupled to a 
continuum of states. Excitation is done by some additional mechanism, 
included in the Hamiltonian of the system $H$. We start with a master 
equation of type (\ref{masterlind}), and in this model
with a single Lindblad operator $L$,
\begin{equation}\label{masterlind2}
\dot{\rho}=\mathcal{L}\rho=-\I [ H,\rho ] 
+ \frac{1}{2}( [ L \rho , L^\dagger ]
+ [ L , \rho L^\dagger ] ).
\end{equation}
In the following we abbreviate the right hand side of the equation
with the superoperator $\mathcal{L}\rho$.
For concreteness, consider $L$ to describe a spontaneous transition from 
level $|1\rangle$
to level $|0\rangle$ with rate $W$, i.e.\ $L=\sqrt{W}|0\rangle\langle 1|$.
We introduce the superoperator
$\mathcal{S}$ such that
\begin{eqnarray*}
\mathcal{H}_{\rm eff}=(\mathcal{L}-\mathcal{S})\rho&=&-\I [ H,\rho ] 
- \frac{1}{2}( L^\dagger L \rho
+ \rho L^\dagger L )\\ \mathcal{S}\rho &=&L\rho L^\dagger.
\end{eqnarray*}
The latter is referred to as the {\it jump} operator since its
describes an emission process accompanied by the replacement
of the density operator $\rho$ with
the ground state: 
$L\rho L^\dagger = W\langle 1|\rho|1\rangle \;|0\rangle\langle 0|$.
With $\mathcal{S}$ such defined,
one obtains the {\it quantum jump representation} \cite{Carmichael} 
of the solution of (\ref{masterlind2}) in the form
\begin{eqnarray}\label{quantumjump}
\lefteqn{\rho (t)= \sum_{m=0}^\infty \int_0^t dt_m \int_0^{t_m} dt_{m-1}
\cdots \int_0^{t_2} dt_1} \qquad \qquad \qquad \\ \nonumber
& &
\times \underbrace{e^{\mathcal{H}_{\rm eff}(t-t_m)} \mathcal{S}
e^{\mathcal{H}_{\rm eff}(t_m-t_{m-1})}
\mathcal{S}\cdots\mathcal{S} e^{\mathcal{H}_{\rm eff}t_1}
\rho(0)}_{\bar{\rho}_c(t)}.
\end{eqnarray}
Clearly, the solution $\rho(t)$ is a sum (or integral, respectively) 
over any number $m$ of emission processes (number of projections 
onto $|0\rangle\langle 0|$
due to the application of the jump operator $\mathcal{S}$), appearing at any
times $t_1, t_2,\ldots,t_m$ between zero and the current
time $t$. One has to integrate over all corresponding (unnormalized)
density operators $\bar{\rho}_c (t)$, as apparent from expression 
(\ref{quantumjump}). Thus, one particular {\it quantum trajectory}
is the normalized density operator
$\rho_c(t)=\bar{\rho}_c(t)/tr \{ \bar{\rho}_c(t) \}$
which describes the time evolution of the quantum
system {\it conditioned} on the particular measurement record,
i.e.\ conditioned on the number and times of emission processes. 
The quantum trajectory 
${\rho}_c(t)$ occurs with probability $tr \{ \bar{\rho}_c(t) \}$.

The normalized quantum trajectory
$\rho_c(t)$ may be determined directly
through the following prescription:
at time $t+\Delta t$ the new density operator $\rho_c(t+\Delta t)$
is obtained in one of two ways:

First, the probability $P_{\rm jump}$, to undergo
a quantum jump, i.e.\ to emit
a quantum during the time interval $\Delta t$ is equal to the 
jump rate times
the length of the time interval times the probability to be in the
excited state: $P_{\rm jump} = W\langle 1|\rho_c(t)|1\rangle \Delta t
= $tr$(L^\dagger L\rho_c(t))\Delta t=$tr$\{ \mathcal{S} \rho_c(t) \} \Delta t$.
If a quantum is emitted (and thus detected), i.e.\ a jump has occurred,
the conditioned quantum state is the ground state: 
$\rho_c(t+\Delta t) = \rho_{\rm jump} = |0\rangle\langle 0| = 
\mathcal{S}\rho_c(t)/tr \{ \mathcal{S} \rho_c(t) \}$. If however,
no jump occurs, the new density operator is given by
\begin{equation}
\rho_c(t+\Delta t)= \rho_{\rm no jump} =
\frac{e^{\mathcal{H}_{\rm eff}\Delta t}\rho_c(t)}
{tr \{ e^{\mathcal{H}_{\rm eff}\Delta t}\rho_c(t) \}}
\end{equation}
as is apparent from the representation (\ref{quantumjump}).
In practice therefore, a quantum trajectory is obtained by determining
a random number $r$ between zero and one in each time step $\Delta t$:
if $r\le P_{\rm jump}$, we set
$\rho_c(t+\Delta t)= \rho_{\rm jump}$, if, however
$r>P_{\rm jump}$, we set
$\rho_c(t+\Delta t)= \rho_{\rm no jump}$.
The full ensemble of possible states is thus given by
$\rho(t+\Delta t) =  P_{\rm jump} \rho_{\rm jump} 
+ (1- P_{\rm jump}) \rho_{\rm no jump}$ and indeed, one may easily
verify that the right hand side equals $\mathcal{L}\rho\;\Delta t$
as expected from the master equation (\ref{masterlind2})
for the full ensemble.

This branching may occur at any time step and a thus huge ensemble of
different quantum trajectories may be obtained. As mentioned before,
the usual
reduced density operator is obtained by taking the ensemble mean.
In order to obtain counting statistics as in the following
sections, we simply average over many runs and obtain numerically
a distribution of jump times as in a real experiment involving
a single quantum system.

\section{Counting statistics and state tomography}

An electron spin on a quantum dot has been found useful as a memory device or 
a qubit for quantum information processing. Readout of the spin state
through a tunneling current was investigated using a rather restricted 
parameter regime for which analytical results were obtained in Ref.\ \onlinecite{Engel2}.

First we want to show, how the analytical results emerge very easily and 
directly from the quantum trajectory approach. Now we consider a regime 
where we can neglect spin flips, i.e.\ $\Wud=\Wdu=\Delta_x=0$. As in 
Ref.\ \onlinecite{Engel2} we choose spin-polarized leads, 
$\gamma^\downarrow=0=\Wsu=\Wus$, and $\gamma^\uparrow=W$. In the limit 
$kT\ll \Delta_z$ we then have $\Wsd=\Wds=W$. The initial state is 
$\spindown$ and since no spin flip occurs on the time scale of interest 
the only processes that happen are transitions between $\spindown$ and 
$\singlet$. The {\it quantum jump representation} (\ref{quantumjump}) 
of this particular solution then reads
\begin{eqnarray}\label{quantumjump2}
\lefteqn{\rho (t)= \sum_{m=0}^\infty \int_0^t dt_m \int_0^{t_m} dt_{m-1}
\cdots \int_0^{t_2} dt_1} \qquad \qquad \qquad \\ \nonumber
& &
\times e^{\mathcal{H}_{\rm eff}(t-t_m)} \mathcal{S}_{ij}
e^{\mathcal{H}_{\rm eff}(t_m-t_{m-1})}
\mathcal{S}_{ji}\cdots\mathcal{S}_{S\downarrow} e^{\mathcal{H}_{\rm eff}t_1}
\rho(0)
\end{eqnarray}
with $i,j=\downarrow,S$ and 
$\mathcal{S}_{ij}\rho=W|i\rangle \langle j|\rho |j\rangle \langle i|$. In the 
regime chosen we can write 
$e^{\mathcal{H}_{\rm eff}(t_k-t_{k-1})}\bar{\rho}_c(t_{k-1})=
e^{-W(t_k-t_{k-1})}\bar{\rho}_c(t_{k-1})$ and then get
\begin{eqnarray}\label{quantumjump3}
\rho(t)&=&e^{-Wt}\sum_{m=0}^\infty \int_0^t dt_m \int_0^{t_m} dt_{m-1}\cdots \int_0^{t_2} dt_1 \mathcal{S}_{ij} \mathcal{S}_{ji} \cdots \mathcal{S}_{S\downarrow} \rho(0)
\\
&=&e^{-Wt}\sum_{m=0}^\infty \frac{(Wt)^m}{m!}|i\rangle 
\underbrace{\langle j|\rho(0)|j\rangle}_{=1} \langle i|.
\end{eqnarray}
Here every operator $\mathcal{S}_{S\downarrow}$ describes an electron 
tunneling onto the dot from lead one and $\mathcal{S}_{\downarrow S}$ 
represents the hopping of an electron from the dot into lead two. Since the 
initial state is $\spindown$, the first transition is 
$\spindown \rightarrow \singlet$ and with a second transition back 
to $\spindown$ the first electron accumulates in lead two. A third transition 
to the singlet does not change the number of electrons in lead two. For a 
particular $q$ (number of electrons in lead two) we have to consider 
$m=2q$ and $m=2q+1$ and the (unnormalized) density operator for a certain 
$q$ at time $t$ is
\begin{equation}\label{quantumjump4}
\rho(q,t)=e^{-Wt}\sum_{m=2q}^{2q+1} \frac{(Wt)^m}{m!}|i\rangle \langle i|.
\end{equation}
Therefore, the probability to find exactly $q$ electrons in lead two at 
time $t$ is
\begin{equation}\label{quantumjump5}
P(q,t)={\rm Tr}\{\rho(q,t)\}=e^{-Wt}\frac{(Wt)^{2q}}{2q!}
\left( 1+\frac{Wt}{2q+1} \right),
\end{equation}
confirming the findings in Ref.\ \onlinecite{Engel2}.

With the general quantum jump representation (\ref{quantumjump}), we
can overcome the limitations of the analytical result, considering
arbitrary regimes and investigating the dynamics numerically. 
So far, the proposed measurement scheme allows one to deduce the 
probability to be in either of the two spin states from the current 
through the dot. A relative phase between $\spinup$ and
$\spindown$, however, cannot be detected. In order
to measure the full spin state, therefore, a tomographical measurement
setup is required. Here, the freedom to apply the ESR field comes into
play. We show that while applying an ESR field, phase-sensitive 
counting statistics result, leading to clear identification of the
qubit state on the Bloch sphere. As in quantum optical setups,
the full state could also be obtained with appropriate $\pi/2$-pulses, 
that effectively change the measurement axis. In this way,
not only the $\langle\sigma_z\rangle$-component as in the original 
proposal, but also
$\langle\sigma_x\rangle$ and $\langle\sigma_y\rangle$ and thus the full $\rho$
can be measured. A simpler concept, not involving these precise
pulses, is to measure the spin state via counting statistics of 
a current through the dot in conjunction with a constant ESR field as we will 
show in the following. For this scheme to be successful it is crucial to 
control the interaction between dot and leads. We are not interested in the
asymptotic, stationary distribution, but in the typical time between
switching the coupling on and the first (or second, or third and so on)
electron appearing in lead two. Also, it is not necessary to
be able to measure the electrons in lead two with a high temporal
resolution: one can switch off the coupling between dot and lead two
after a certain time $t$ and has any time thereafter to collect the
electrons in lead two. We note that different measurement schemes are 
possible. Since an electron tunneling onto the dot already carries the 
information about the spin of the dot electron, one could 
abandon lead two altogether and try to
monitor the number of electrons on the quantum dot, 
e.g.\ with a quantum point contact. Our proposal for quantum state 
tomography could be transferred to other setups as well, as for
recent experiments.\cite{Elzerman3, Jelezko}

We assume that the dot is in a given initial state at $t=0$, when the
coupling to the leads is switched on. Then we measure the
number of electrons tunneling 
into lead two. According to the quantum trajectory approach we calculate the 
evolution of the density matrix. Every jump from $\singlet$ to 
$\spindown$ or $\spinup$ indicates that an electron tunneled out of the dot.
At very low temperatures, as assumed throughout this paper,
the probability of tunneling into lead two is close to unity, 
while tunneling into lead one is very unlikely. 

A single run of the stochastic evolution will display emission
processes, i.e.\ contributions to the current, at certain random times.
Counting the corresponding number of quanta in lead two as a function of
time for a large ensemble of quantum trajectories allows us to determine
the probability $P(q,t)$ of 
finding exactly $q$ electrons in lead two at time $t$ for a given initial 
state of the dot. Such counting distributions are displayed in the following
Figures. We stress again that experimentally, it is not
required to be able to measure these arrival times with a high resolution.
One simply switches off the coupling between dot and lead two after a given 
time $t$. Then one
has any time to determine the number of electrons in lead two.
Our numerical procedure can be applied to any
parameter values and any time dependence of the driving ESR field.
For the regime chosen in Ref.\ \onlinecite{Engel2}, we recover the
analytical results (\ref{quantumjump5})
to a high degree of precision, as will be shown
below.

Let us now turn to the probability distributions $P(q,t)$ of
finding exactly $q$ electrons in lead two at time $t$ for a given initial 
state $\rho$. As we will show, by employing the ESR field, the
counting statistics allows to clearly identify the full two-level state,
including the relative phase.
As usual, we choose to parametrise the latter
through the coordinates on the Bloch sphere: the spin up state $\spinup$
corresponds to the north pole with $r=1,\theta=0^\circ$, while
the spin down state $\spindown$ has coordinates
$r=1,\theta=180^\circ$. The full mixture 
$\rho_0 = \frac{1}{2}\left(\spinup\spinupd + \spindown\spindownd\right)$
corresponds to the center of the Bloch sphere,
$r=0$ while coherent superpositions 
$\psi = (\spinup + e^{i\phi}\spindown)/\sqrt{2}$ live on the
equator with $r=1,\theta = 90^\circ, \phi$.

\begin{figure}
\includegraphics[angle=0,width=0.5\textwidth]{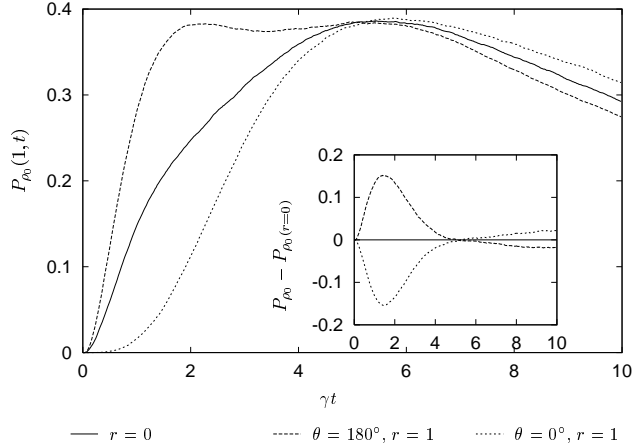}
\caption{\label{downupOSP}
Counting statistics $P(q=1,t)$ for the first electron
with initial spin up state (dashed ,$r=1,\theta = 0$),
spin down state (dotted, $r=1,\theta = 180^\circ$) and the totally 
mixed state (full curve, $r=0$). 
The respective coordinates refer to the Bloch sphere. The inset shows
the same curves with the counting statistics of the fully mixed
state subtracted. Parameters chosen are
$\Delta_x=2\gamma=4(\Delta_z-\omega)=2\times 10^6 s^{-1}, T=20mK, B_z=12T$}
\end{figure}

In order to be able to use these counting statistics as a method
for spin state tomography, the right choice of parameters is crucial.
From Eqs.\ (\ref{rwa1},\ref{rwa2},\ref{rwa3},\ref{rwa4}) it is obvious 
that coherences
in the two-level state can only be transferred to measurable 
probabilities through
the coupling introduced by the ESR field of magnitude $\Delta_x$.
On the other hand, a large value of $\Delta_x$ leads to
Rabi oscillations and thus prevents us to distinguish clearly the two
fundamental $\spinup$ and $\spindown$ states on a time scale large compared with the Rabi frequency $\Delta_x /2$.
Closer inspection of Eqs.\ (\ref{rwa1}, \ref{rwa2}, \ref{rwa4}) 
and numerical evidence shows that a good phase sensitivity with
preserved distinguishability of $\spinup$ and $\spindown$ is achieved
through the choices
\begin{eqnarray}\label{readoutconditions}
\frac{\Delta_x}{2} & \approx & \Wsd \\ \nonumber
\Delta_z-\omega & \approx & \frac{\Delta_x}{4}.
\end{eqnarray}
Physically, the first condition (on the ESR field strength) means that the 
spin should not be 
flipped to fast (compared with the measurement time scale $\Wsd$) but still,
the ESR field had time enough to make the coherences felt.
The second condition (on the ESR field frequency) ensures
that the method is sensitive to all values of the 
phase angle $\phi$.

In Fig.\ \ref{downupOSP} we show counting statistics for the first
electron $P(q=1,t)$ to appear in lead two. 
We choose the transition rate $\gamma=10^6 s^{-1}$, an experimentally 
accessible
magnetic field strength\cite{lossprivate} of the ESR
field $B_x \approx 5.16 G$, a slightly detuned ESR field frequency 
$\Delta_z-\omega=5\times 10^5 s^{-1}$, a temperature $T=20mK$,
and a static magnetic field of strength $B_z=12T$.
For the ESR field to start at zero we choose the fixed phase 
$\varphi=3\pi/2$. 
Furthermore, we assume $T_1=10^{-4}s$ and $T_2=10^{-5}s$
for the intrinsic relaxation and decoherence times.
All Figures are calculated with an ensemble of 50000 trajectories.

The spin down state only allows for electrons to tunnel through the
dot, which is clearly visible in the counting statistics: if the spin
starts off in the spin up state (dotted curve) the time to measure
the first electron is delayed compared to the mixture and even
more so compared to the spin down state. Eventually, however, due to
the presence of the ESR field, a sufficient spin down component will
be established allowing electrons to tunnel through the dot. Still, 
both states are clearly distinguishable through their counting 
statistics.

\begin{figure}
\includegraphics[angle=0,width=0.5\textwidth]{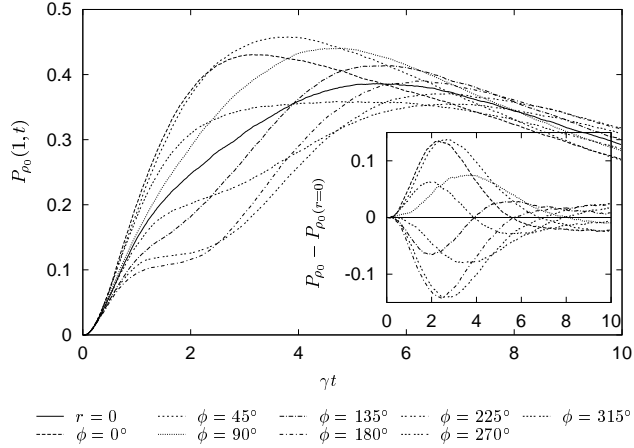}
\caption{\label{equatorOSP}
Counting statistics $P(q=1,t)$ for eight coherent superpositions
$\psi = (\spinup + e^{i\phi}\spindown)/\sqrt{2}$
along the equator of the Bloch sphere ($r=1,\theta=90^\circ$, various
angles $\phi$) and the fully mixed state ($r=0$, full line). 
The inset shows the same curves with $P(q=1,t)$ of the full mixture 
subtracted. Same parameters as in Fig.\ \ref{downupOSP}.}
\end{figure}

Not only are counting statistics useful to distinguish
between up and down state. The arrival time distribution also 
differentiates between coherent superpositions and mixtures. 
In conjunction with the 
ESR field one may even determine the phase of coherent superpositions
of type $\psi = (\spinup + e^{-i\phi}\spindown)/\sqrt{2}$
as displayed in Fig.\ \ref{downupOSP}. The full line corresponds
to $P(q=1,t)$ of a fully mixed initial state ($r=0$), the dashed and
dotted lines correspond to eight coherent superpositions along the
equator of the Bloch sphere. Clearly, $P(q=1,t)$ shows different
behavior for different angles $\phi$ and may thus be used to
fully identify the initial state.

As we have seen, with these choices for the ESR field, not only we can
distinguish $\spinup$ from $\spindown$ through counting statistics
as in Fig.\ \ref{downupOSP}. We are in a position to fully determine
the two-level state -- in particular, it is possible to clearly
distinguish a coherent superposition of 
$\spinup$ from $\spindown$ from the mixture of the two, as
shown in Fig.\ \ref{equatorOSP}. 

\begin{figure}
\includegraphics[angle=0,width=0.8\textwidth]{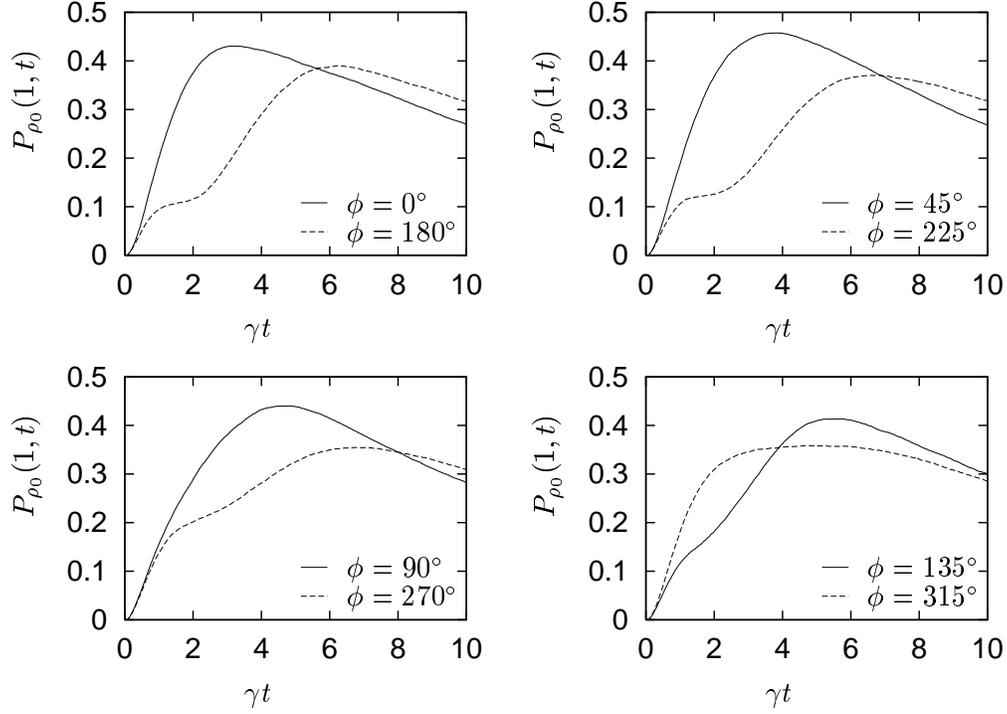}
\caption{\label{diff_paarweise0}
Graphs taken from Fig.\ \ref{equatorOSP} for four pairs of opposite
states along the equator of the Bloch sphere.}
\end{figure}

\begin{figure}
\includegraphics[angle=0,width=0.8\textwidth]
{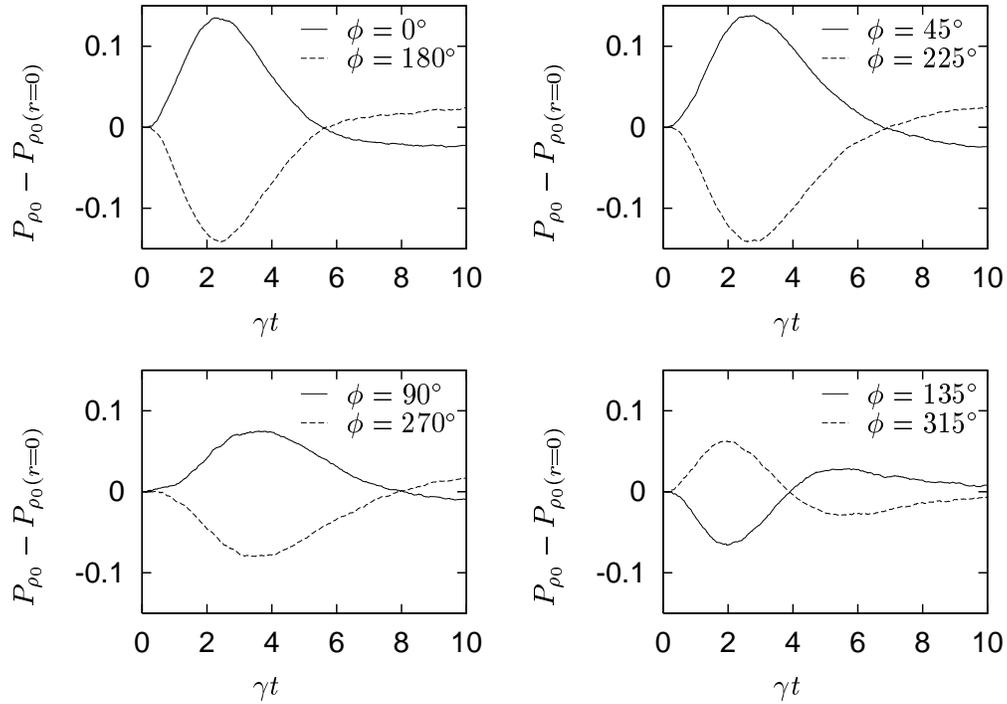}
\caption{\label{diff_paarweise}
Same as Fig.\ \ref{diff_paarweise0} with $P(q=1,t)$ of the
full mixture subtracted. We clearly see the symmetry of the curves for
opposite states on the Bloch sphere.}
\end{figure}

The insets of Figs.\ \ref{downupOSP} and
\ref{equatorOSP} reveal an interesting structure 
underlying the shapes of $P(q=1,t)$:
Once the counting distribution of the full mixture ($r=0$) is
subtracted, statistics of states corresponding to opposite points on 
the Bloch sphere appear as mirror-images of each other, as
highlighted in Figs.\ \ref{diff_paarweise0} and \ref{diff_paarweise}.
In these Figures we display the counting statistics $P(q=1,t)$ for
four pairs of opposite initial states along the equator of the
Bloch sphere and clearly confirm the observations just mentioned.
A linear combination of initial states leads to a linear combination of counting statistics in the ensemble and thus to this symmetry.
Still, each curve in itself seems complicated enough to underline the
importance of our numerical approach. Using the quantum trajectory
method, any time dependence of the fields and any choice of
parameters is possible.

The more mixed the initial state, i.e.\ the smaller $r<1$ on the Bloch
sphere, the closer the curve to the curve of the fully mixed state.
It is also worth noting that we keep the initial phase $\varphi$ of
the ESR pulse fixed for all calculations. An average over all possible
phases would indeed lead to the graph of the fully mixed state, 
irrespective of the phase $\phi$ of the initial quantum state.

\subsection{Higher order statistics and $q=0$}

\begin{figure}
\includegraphics[angle=0,width=0.6\textwidth]{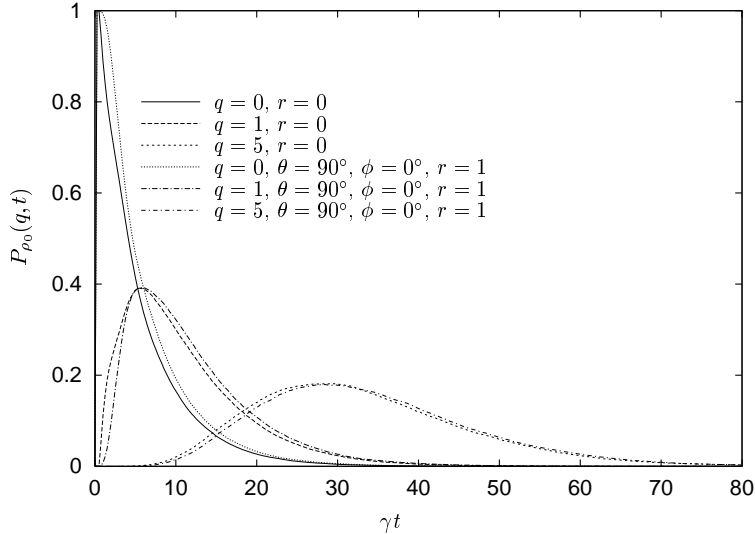}
\caption{\label{qdiagram}
Counting statistics of exactly zero, one, five electrons tunneling through 
the dot. Evidently, if less pronounced, all counting statistics 
$P(q,t)\;(q=0,1,\ldots,5,\ldots)$ carry information about the 
initial quantum state.}
\end{figure}

We close this section by pointing out that also higher order counting 
statistics ($q=2,3,4,5$) display state-sensitive behaviour -- if only less 
pronounced. This is quite obvious since a delayed first tunneling event 
shifts the starting time for the following electrons. For $q=0$, the 
difference between the counting statistics for various initial states
is well pronounced. In this latter case, however, the curves do not cross
which diminishes the distinguishability of states and the best choice 
for that is $q=1$. As displayed in Fig.\ \ref{qdiagram}, higher-order 
counting statistics $P(q,t)$ still distinguishes between the fully mixed 
state ($r=0$) and a coherent superposition ($r=1, \theta=90^\circ$).

\subsection{The role of spin-polarized leads}

\begin{figure}
\includegraphics[angle=0,width=0.5\textwidth]{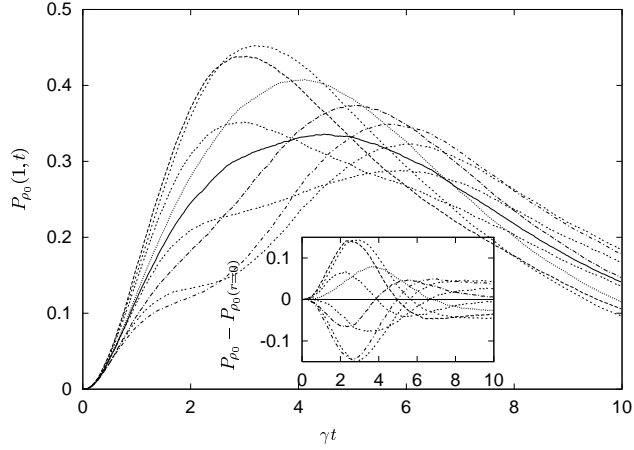}
\caption{\label{equatorMSP}
Same graph as Fig.\ \ref{equatorOSP}. Here, however using
spin-polarized leads with $\gamma^\uparrow=1\times 10^6 s^{-1}$,
$\gamma^\downarrow=0$.}
\end{figure}

The original proposal for the spin state readout was based on
spin-polarized leads in order to clearly distinguish the
two states $\spinup$, $\spindown$ by a single shot measurement. As the 
counting statistics require an ensemble measurement, our results suggest 
that spin polarization is not required for those -- not even advantageous, 
in fact.
In Fig.\ \ref{equatorMSP} we display counting statistics $P(q=1,t)$
for spin-polarized leads (only spin-up electrons in the leads, i.e.\ $\gamma^\downarrow=0$). We notice only marginal differences compared to 
the case of unpolarized leads (Fig.\ \ref{equatorOSP}). 

For large times, it is more likely to observe precisely one electron
in the case of unpolarized leads. The reason for this behaviour is the 
fact that for unpolarized leads, there is also the possibility that
the spin-up electron on the dot (rather than the spin-down electron 
entering the dot) may tunnel out of the dot. Then the dot is in the
ground state and therefore closed for the tunneling of another electron.
It is only after the ESR field had time to populate the excited
state that a second electron may tunnel through the dot. In fact, 
it turns out that this
mechanism is the preferred tunneling event: for the parameters
of Fig.\ \ref{downupOSP} we find $\Wds=1\times10^6 s^{-1}$ and
$\Wus=2\times10^6 s^{-1}$.

\section{Relaxation and decoherence times}

\begin{figure}
\includegraphics[angle=0,width=0.9\textwidth]{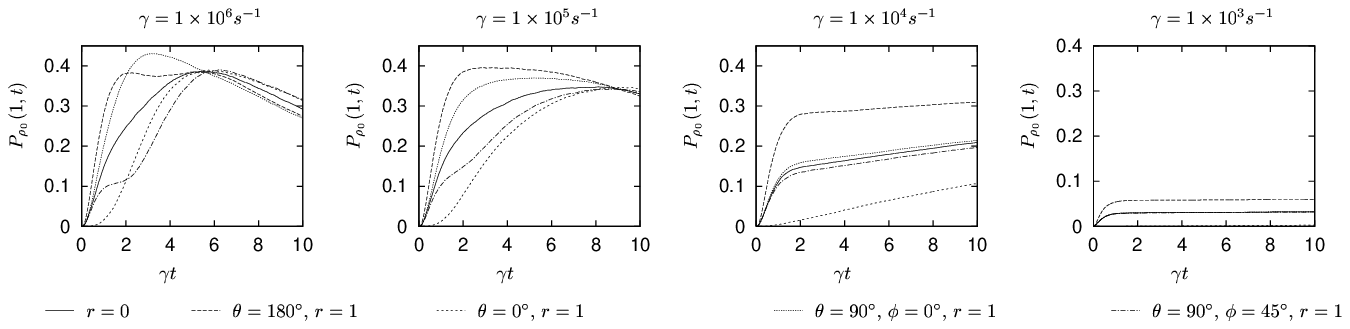}
\caption{\label{Wscan}
Measurement of a variety of states ($\spinup$, $\spindown$, coherent superposition) facilitates an estimate of $T_2$ and $T_1$.}
\end{figure}

Our proposed setup including the ESR field may be used to determine
the intrinsic relaxation time $T_1$ and decoherence time $T_2$ of the 
qubit in the time domain. Tuning the tunneling rate over a wide
range (and adjusting the ESR field strength and frequency according
to conditions (\ref{readoutconditions}), one can easily see the effect
of decoherence and relaxation. In the series of graphs in Fig.\ \ref{Wscan}
we show counting statistics for the $\spindown$ and $\spinup$-state,
for the full mixture and for two coherent superpositions
(states on the equator of the Bloch sphere). Clearly, for large
tunneling rate (left graph, (a)), all states may be distinguished.
The third graph (c) shows a regime where decoherence has fully set in:
while the states  $\spindown$ and $\spinup$-state remain essentially
unaffected, the counting statistics of the coherent superpositions 
collapse onto the curve of the full mixture. In other words,
while no relaxation has set in yet, coherences between the states
$\spindown$ and $\spinup$ have disappeared. Decreasing the
tunneling rate even further, counting statistics finally reveal the
relaxation time: eventually, the initial states
$\spindown$ and $\spinup$ may no longer be distinguished, i.e.\ the relaxation has taken place.

\section{Single shot readout}

Counting statistics was used in
Ref.\ \onlinecite{Engel2} to determine the measurement time and measurement
efficiency. It was shown that after about ten times the tunneling time,
the spin state on the dot could be determined to be either
$\spinup$ or $\spindown$ with close to 100\% efficiency,
even if only a single measurement is made
(provided it was either $\spinup$ or $\spindown$).
Crucially, these results are based on spin-polarized leads.
Without spin-polarization the determination of the dot state
with a single measurement appears problematic. The reason is
that an electron may tunnel from the quantum dot back into lead
one rather than into lead two. In fact, since in the case of
unpolarized leads there are three ways for an electron
to leave the dot:
it may tunnel from lead one onto the dot and further into lead two
(transition $\singlet \rightarrow \spindown$) or
the electrons interchange the roles and the one residing on the
dot tunnels into either of the two leads
($\singlet\rightarrow\spinup$). The latter two possibilities are
almost of same probability and therefore with 
a probability of about one third a tunneling process has taken
place without having been observed in lead two. In other words,
with a probability of two thirds only we can claim that without
detecting an electron in lead two after sufficient time the
spin state on the dot was $\spinup$.

In order to overcome this problem we propose to measure the number
of electrons on the dot with a quantum point contact and recent 
experiments \cite{Elzerman3} show that such a concept may work. If an 
electron tunnels onto the dot it confirms that the state was $\spindown$ 
and if it stays there sufficiently long the QPC as electrometer 
recognizes the charge. Since the projection takes place if an electron 
tunnels onto the dot, the second lead could even be omitted and then the
dot electron would tunnel into lead one. For clarity in the following
argument, however, we do not change our setup and leave lead two as before.

With the help of the QPC it is possible to measure the state of
a single electron spin with an efficiency of almost 100\% after about
ten times the tunneling time, as for polarized leads as shown
in Fig.\ \ref{proj_nc}. Note that for this scheme to work, it is important
that the time resolution of the point contact measurement has to be 
better than the inverse tunneling rate $\gamma^{-1}$
to ensure the detection of the tunneled electron on the dot. 

\begin{figure}
\includegraphics[angle=0,width=0.5\textwidth]{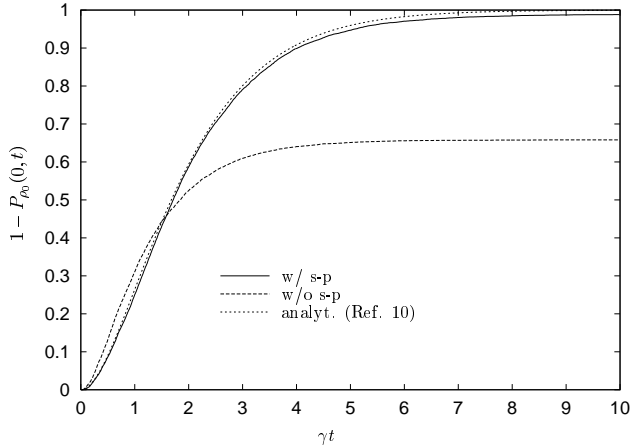}
\caption{\label{proj}
With spin-polarized (s-p) leads, a detection of an electron
in lead two after about 10 times the tunneling time ensures to
almost 100\% that the initial dot state was $\spindown$. Without spin
polarization, only a probability of about two thirds may be achieved
due to the possibility of the electron to tunnel into lead one. The numerical curve accounts for spin relaxation and therefore differs slightly from the analytical result. Without spin relaxation no difference can be seen.
Parameters as in Fig.\ \ref{downupOSP} respectively Fig.\ \ref{equatorMSP}, but without ESR-field.}
\end{figure}

\begin{figure}
\includegraphics[angle=0,width=0.5\textwidth]{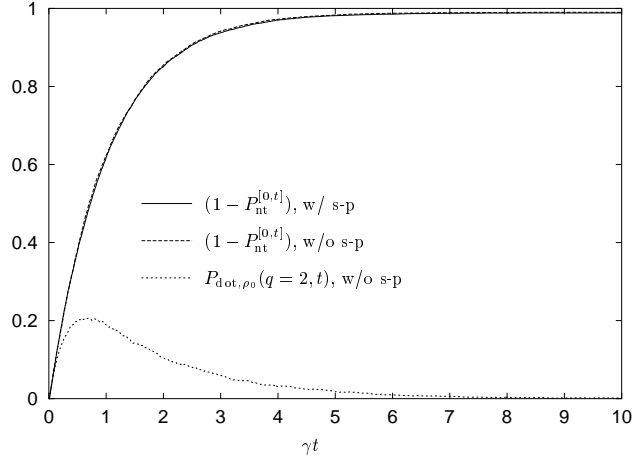}
\caption{\label{proj_nc}
Measuring the probability to find two electrons on the dot
one may determine the dot state with 100\% efficiency. Clearly, the
time resolution has to be better than the typical time the two 
electrons stay on the dot, i.e.\ better than $\gamma^{-1}$. 
Here $P_{{\rm nt}}^{[0,t]}$ is the probability that no tunneling event occurred in the whole interval $[0,t]$ and $P_{{\rm dot},\rho_0}(q,t)$ is the probability to find exactly $q$ electrons on the dot at time $t$. The initial state was always $\spindown$, parameters as in Fig.\ \ref{proj}.}
\end{figure}

\section{Conclusion}
We use quantum trajectory methods to investigate counting statistics
of electrons tunneling through a quantum dot. We show how an additional
ESR field may actively be used to perform a full ``state tomography''.
Applying the field during the measurement allows one to clearly
identify the coherences between the two superposed states. We discuss
the relevance of our findings for determining intrinsic relaxation
and decoherence times of electron spin states in quantum dots --
in the time domain. Based on a quantum point contact we propose a scheme for single-shot readout without the need for spin-polarized leads. Similarities of the investigated quantum dot to three-level systems in quantum optics (e.g.\ so-called ``V''-systems) are evident. We underline these
connections by applying the
quantum jump method in order to {\it unravel} the dynamics of the
full density operator $\rho(t)$ into
subensembles $\rho_c(t)$ corresponding to certain measurement records
in the leads. Thus, we describe the {\it conditioned} time evolution
of the spin state, given a certain measurement record, as in actual
experiments with single quanta.

We believe that such connections between methods of quantum optics
and mesoscopic devices will prove more and more useful in the future
as nanotechnology achieves further breakthroughs in the coherent
manipulation of quantum dynamics in solid state devices.

\begin{acknowledgments}
We thank D. Loss, H. A. Engel, and J. M. Elzerman for useful discussions.
\end{acknowledgments}

\end{document}